\documentclass[journal]{IEEEtran}
\usepackage{tabularx}
\usepackage{longtable}
\usepackage{subfigure}
\usepackage{enumitem}
\usepackage{amsmath,amssymb,amsthm}
\usepackage{mathtools}
\usepackage{graphicx}
\usepackage{cite}
\usepackage{url}
\usepackage{algorithm}
\usepackage{algpseudocode}
\usepackage{booktabs}
\usepackage{tabularx}
\usepackage{booktabs}

\newtheorem{definition}{Definition}

\begin{document}

\title{Safety-Critical Reinforcement Learning with Viability-Based Action Shielding for Hypersonic Longitudinal Flight}

\author{Hossein~Rastgoftar%
\thanks{Hossein Rastgoftar is with the Department of Aerospace and Mechanical Engineering,
The University of Arizona, Tucson, AZ 85719, USA. {\tt\small hrastgoftar@arizona.edu}}%
}

\maketitle

\begin{abstract}
This paper presents a safety-critical reinforcement learning framework for nonlinear dynamical
systems with continuous state and input spaces operating under explicit physical constraints. 
Hard safety constraints are enforced independently of the reward through action shielding and
reachability-based admissible action sets, ensuring that unsafe behaviors are never intentionally
selected during learning or execution. To capture nominal operation and recovery behavior within a
single control architecture, the state space is partitioned into safe and unsafe regions based on
membership in a safety box, and a mode-dependent reward is used to promote accurate tracking inside
the safe region and recovery toward it when operating outside. To enable online tabular learning on
continuous dynamics, a finite-state abstraction is constructed via state aggregation, and action
selection and value updates are consistently restricted to admissible actions. The framework is
demonstrated on a longitudinal point-mass hypersonic vehicle model with aerodynamic and propulsion
couplings, using angle of attack and throttle as control inputs.

\end{abstract}

\section{Introduction}
In safety-critical autonomous systems operating under stringent aerodynamic, thermal, and actuator
constraints, the direct application of reinforcement learning (RL) remains challenging due to the risks
associated with unconstrained exploration and the difficulty of encoding safety requirements purely
through reward design. In high-consequence regimes, reward functions alone are often insufficient to
prevent unsafe behavior during learning, while exploratory actions that violate physical or
operational constraints can lead to irrecoverable states or catastrophic failure. These challenges
motivate RL frameworks that enforce hard safety constraints independently of the
reward, restrict exploration to admissible behaviors, and explicitly account for the state-dependent
structure of safety-critical operation. Such considerations are particularly important for
high-speed aerospace vehicles, where nonlinear dynamics, tight operating envelopes, and limited
tolerance for failure demand safety preservation throughout both learning and deployment.

\subsection{Related Work}

Hypersonic flight dynamics have been widely studied using control-oriented longitudinal models that capture the strong coupling among aerodynamics, propulsion, atmospheric effects, and altitude-dependent gravity, while remaining tractable for control design and simulation. A nonlinear, physics-based longitudinal model incorporating aerodynamics, scramjet propulsion, and structural flexibility is developed in \cite{BolenderDoman2007JSR}. This framework is subsequently simplified into a control-oriented form in \cite{ParkerBolenderDoman2007JGCD} to facilitate nonlinear control analysis and synthesis. A comprehensive survey of hypersonic vehicle dynamics and control methodologies, highlighting challenges such as flexibility, strong coupling, and nonminimum-phase behavior, is provided in \cite{Xu2015OverviewHypersonic}. Trajectory optimization for hypersonic glide vehicles with peak normal load as the objective is addressed in \cite{Wang2019AST} using mixed-integer and sequential convex optimization techniques.

RL has seen increasing application in aerospace control due to its ability to address nonlinear dynamics and model uncertainty \cite{6315769,waslander2009wind,11073806}. A rigorous connection between RL and feedback control is established in \cite{6315769}, where actor–critic methods are shown to learn optimal adaptive controllers online by effectively solving Hamilton–Jacobi–Bellman equations in real time without requiring full knowledge of system dynamics. Building on classical guidance laws, \cite{11073806} presents a deep Q-network–based adaptive proportional navigation scheme that learns online pitch- and yaw-gain adjustments on top of standard PNG, yielding significant improvements in three-dimensional interception performance against highly maneuvering targets while preserving interpretability and energy efficiency. Koch et al. demonstrate that deep RL—particularly PPO—can achieve high-precision UAV inner-loop attitude control when trained in a high-fidelity digital-twin environment (GymFC), matching or surpassing well-tuned PID controllers while retaining adaptability to uncertain dynamics \cite{koch2019reinforcement}.

Recent studies have extended RL-based adaptive control and trajectory optimization to hypersonic vehicles operating under severe aerodynamic uncertainty \cite{SciOpenRLHypersonic,SciOpenHypersonicTrajectory2025,gaudet2022adaptive,gaudet2023deep}. Gaudet et al. develop a reinforcement meta-learning–based adaptive guidance law for hypersonic gliders that combines recurrent PPO with curved-space parallel navigation, enabling robust adaptation to aerodynamic uncertainty and actuator or sensor faults and outperforming optimal-trajectory LQR tracking in off-nominal conditions \cite{gaudet2022adaptive}. In related work, \cite{gaudet2023deep} proposes a CNN-based deep RL policy for real-time weapons-to-target assignment in hypersonic strike scenarios, achieving near-optimal performance relative to nonlinear integer programming while reducing computation time by orders of magnitude. Sethi introduces an RL-augmented adaptive control framework that integrates a baseline model-based controller with online RL to compensate for aerodynamic uncertainties and disturbances, improving trajectory tracking and stability without requiring accurate system models \cite{SciOpenRLHypersonic}.

Despite these advances, most existing RL-based aerospace control approaches enforce safety constraints only indirectly through reward shaping or episode termination, which can lead to unsafe exploration or degraded constraint satisfaction in safety-critical flight regimes. This limitation has motivated significant interest in safe RL, with surveys reviewing constrained Markov decision processes (MDP), Lyapunov-based methods, and risk-sensitive formulations that enforce constraints in expectation or with high probability \cite{GarciaFernandez2015SafeRL,SafeRL_Survey2025}. Constrained policy optimization and related approaches explicitly incorporate constraints into policy updates \cite{Achiam2017CPO,Chow2019LyapunovSafePO}, but often depend on function approximation, dual optimization, or accurate constraint gradients, complicating deployment in high-fidelity physical systems. 

An alternative and complementary approach to safety is shielding, in which unsafe actions are filtered online using model-based prediction or formal verification, providing strong safety guarantees without altering the underlying learning algorithm \cite{Alshiekh2018Shielding,ShieldingRL_CACM2025}. Hou et al. present a theoretical and empirical study of invalid action masking in reinforcement learning, proving that the naive policy gradient remains valid and proposing off-policy and composite-objective algorithms that enable agents to learn to avoid invalid actions while retaining robust performance after masking is removed \cite{hou2023exploring}. A unifying decision-theoretic planning framework based on Markov decision processes shows how structural properties such as abstraction, aggregation, and decomposition can be exploited to achieve computationally tractable planning under uncertainty. Kveton et al. develop hybrid approximate linear programming for factored MDPs, providing theoretical error bounds and efficient solvers that avoid discretization in mixed discrete–continuous domains \cite{kveton2006solving}. In safety-critical control, such hybrid formulations naturally represent nominal operation, contingency management, and recovery behaviors within a unified decision-making architecture. Recent advances in safe reinforcement learning leverage this structure: \cite{11312958} introduces an on-policy safe policy iteration method for nonlinear discrete-time systems with input saturation, guaranteeing safe initialization, exploration, and learning through CLF–CBF–based constraints, barrier-augmented rewards, and a fallback safety policy, while \cite{11312525} develops a model-based MPC–RL framework with probabilistic control barrier functions, combining sampling-based stochastic safety guarantees with learned terminal costs and barrier parameters for efficient and probabilistically safe control.


\subsection{Contributions}

This paper presents a safety-critical RL framework for hypersonic longitudinal
flight control that integrates high-fidelity continuous dynamics with a finite abstraction and
explicit safety enforcement.
The approach combines an offline reachability-based feasibility analysis with online tabular
RL under hard physical constraints.
The learning problem is interpreted as a hybrid Markov decision process (MDP) with mode-dependent
objectives induced by safety-set membership and a uniformly enforced admissible action mask.
Rather than proposing a new RL algorithm, the contribution lies in a principled
integration of hybrid reward modeling, action shielding, and mask-consistent tabular learning
tailored to hypersonic flight envelopes. In this context, the main contributions are as follows:

\begin{itemize}
\item \textbf{Mode-dependent reward formulation for constrained flight control:}
Building on prior work on factored and hybrid reward structures \cite{Boutilier1999,Hauskrecht2011}, we formulate hypersonic longitudinal control as a single constrained MDP in which safety-set membership induces nominal and recovery reward modes. The reward structure adapts to safety status, while the system dynamics and admissible action constraints remain unchanged, enabling hybrid reward modeling without introducing mode-dependent dynamics or controller switching.

\item \textbf{Viability-based admissible action synthesis ensuring forward invariance:}
Unlike one-step shielding approaches that filter actions based solely on instantaneous safety predictions \cite{Alshiekh2018,Konighofer2025}, we construct a state-dependent admissible action set through an offline reachability analysis over a discretized abstraction. This fixed-point computation eliminates not only actions that violate hard constraints, but also dynamically admissible actions that lead to non-viable states from which constraint satisfaction cannot be maintained. The resulting admissible action set guarantees forward invariance of the feasible abstract state set and is enforced uniformly during learning and execution.

\item \textbf{Mask-consistent and locally smooth tabular Q-learning under hard action constraints:}
Motivated by recent analyses of optimistic bias arising from inconsistent invalid-action masking \cite{Huang2020,Hou2023}, we develop a tabular Q-learning framework in which both action selection and Bellman backups are explicitly restricted to the state-dependent admissible action set. To address practical control requirements in discretized hypersonic flight, we further incorporate a neighborhood-based local action selection mechanism that prioritizes admissible actions near the previously applied command, improving closed-loop smoothness without modifying the underlying value update or safety guarantees.

\item \textbf{Constraint-aware episode chaining for long-horizon recovery learning:}
We introduce a constraint-aware episode chaining mechanism that propagates terminal states across episodes to expose long-horizon recovery behavior under sustained constraint satisfaction. Chaining is disabled following hard-constraint violations, preventing the propagation of infeasible initial conditions and reinforcing learning within the forward-invariant feasible set.

\item \textbf{Application to hypersonic longitudinal flight dynamics:}
The proposed framework is demonstrated on a hypersonic longitudinal flight model incorporating altitude-dependent gravity, atmospheric density variation, Mach-dependent aerodynamics and propulsion, and aerothermal and structural envelope constraints.
\end{itemize}

\subsection{Outline}
This paper is organized as follows. The scope of the work is explained in Section~\ref{Problem Statement}. A high-fidelity model of hypersonic longitudinal motion is presented in Section~\ref{Flight Dynamics}. The physical and operational constraints of hypersonic flight are formulated as soft and hard constraints in Section~\ref{sec:constraints}. An RL-based hybrid control framework is developed in Section~\ref{HybridFramework} to ensure safe operation during hypersonic flight. Simulation results are presented in Section~\ref{sec:simulation}, followed by concluding remarks in Section~\ref{Conclusion}.

\begin{table*}[t]
\centering
\footnotesize
\caption{Notation used in the hypersonic RL framework.}
\label{tab:nomenclature}
\renewcommand{\arraystretch}{1.05}
\setlength{\tabcolsep}{5pt}
\begin{tabular}{lll}
\toprule
\textbf{Variable} & \textbf{Explanation} & \textbf{Value / Units} \\
\midrule
$x,\,x_k$ & Continuous / discrete system state & $[h,V,\gamma,m]^\top$ \\
$u,\,u_k$ & Continuous / discrete control input & $[\alpha,\delta]^\top$ \\
$\Delta t$ & Sampling period & s \\
$g(\cdot,\cdot)$ & Continuous-time dynamics & $\dot x=g(x,u)$ \\
$f(\cdot,\cdot)$ & Discrete-time dynamics (RK2) & $x_{k+1}=f(x_k,u_k)$ \\
$h$ & Altitude & m \\
$V$ & Velocity magnitude & m/s \\
$\gamma$ & Flight-path angle & rad \\
$m$ & Vehicle mass & kg \\
$\alpha$ & Angle of attack & rad \\
$\delta$ & Throttle command & $[0,1]$ \\
$\rho(h)$ & Atmospheric density & kg/m$^3$ \\
$a(h)$ & Speed of sound & m/s \\
$M$ & Mach number, $V/a(h)$ & -- \\
$S$ & Reference aerodynamic area & m$^2$ \\
$C_L(\alpha,M)$ & Lift coefficient & aerodynamic map \\
$C_D(\alpha,M)$ & Drag coefficient & aerodynamic map \\
$L,\,D$ & Lift and drag forces & N \\
$T$ & Thrust & N \\
$T_{\max}(h,M)$ & Maximum thrust & N \\
$I_{\mathrm{sp}}(h,M)$ & Specific impulse & s \\
$\dot m_f$ & Fuel mass flow rate & kg/s \\
$g(h)$ & Gravity & m/s$^2$ \\
$g_0$ & Standard gravity & 9.80665 m/s$^2$ \\
$q(h,V)$ & Dynamic pressure & Pa \\
$q_{\max}$ & Maximum dynamic pressure & Pa \\
$n_z$ & Normal load factor $L/(m g)$ & -- \\
$n_{z,\max}$ & Maximum load factor & -- \\
$\dot Q(h,V)$ & Aerodynamic heating surrogate & model \\
$\dot Q_{\max}$ & Maximum heating rate & limit \\
$\mathcal{X}$ & Continuous state space & $\subset\mathbb{R}^4$ \\
$\mathcal{B}_{\mathrm{safe}}$ & Safety box in $(h,V,\gamma)$ & definition \eqref{eq:safetybox} \\
$\mathcal{S}$ & Abstract state space & finite \\
$s_k$ & Abstract state & index \\
$\Pi(\cdot)$ & State aggregation operator & binning \\
$\mathcal{A}$ & Discrete action set & finite \\
$a_k$ & Discrete action & element of $\mathcal{A}$ \\
$Q(s,a)$ & Action-value function & tabular \\
$r_k$ & Reward & scalar \\
$\gamma$ & Discount factor & $(0,1)$ \\
$\alpha_k$ & Learning rate & $(0,1)$ \\
$\mathcal{A}_{\mathrm{mask}}(x)$ & Admissible actions (one-step hard-safe) & \eqref{eq:Amask} \\
\bottomrule
\end{tabular}
\end{table*}
\section{Problem Statement}\label{Problem Statement}

This paper addresses the longitudinal flight control of an air-breathing hypersonic vehicle operating under nonlinear, uncertain dynamics and strict safety constraints. The vehicle must maintain around a desired cruise conditions, inside a safety box, while respecting state and actuator limits associated with the hypersonic flight envelope and atmospheric interactions. Model uncertainty, external disturbances, and strong coupling between aerodynamics and propulsion make purely model-based control approaches difficult to tune and potentially brittle in off-nominal conditions.

The objective is to design a hybrid RL–based control architecture that improves longitudinal flight performance through online adaptation while preserving safety at all times. Specifically, the controller should (i) enforce safety-critical constraints during both learning and execution and (ii) reliably transition to recovery behaviors when unsafe conditions are encountered during longitudinal flight control. The resulting framework seeks to combine the adaptability of RL with the robustness and interpretability required for safety-critical hypersonic flight control. 

Table~\ref{tab:nomenclature} summarizes the notation used throughout the paper.

\section{Hypersonic Vehicle Model}\label{Flight Dynamics}

We model the vehicle longitudinal motion in a vertical plane with state
\[
x \triangleq (h,\;V,\;\gamma,\;m)\in \mathbb{R}_{\ge 0}\times\mathbb{R}_{>0}\times\mathbb{R}\times\mathbb{R}_{>0},
\]
where $h$ is altitude (m), $V$ is speed (m/s), $\gamma$ is the flight-path angle (rad), and $m$ is vehicle mass (kg).
The control input is
\[
u \triangleq (\alpha,\;\delta),
\]
where $\alpha$ is angle of attack (rad) and $\delta\in[0,1]$ is throttle. 
The longitudinal equations of motion are given by \cite{Anderson2006Hypersonic}
\begin{subequations}\label{eq:hyp_dyn_ct}
\begin{align}
\dot h &= V\sin\gamma, \label{eq:hyp_hdot}\\
\dot V &= \frac{T\cos\alpha - D - m g(h)\sin\gamma}{m}, \label{eq:hyp_Vdot}\\
\dot\gamma &= \frac{L + T\sin\alpha - m g(h)\cos\gamma}{m\,\max\{V,1\}}, \label{eq:hyp_gdot}\\
\dot m &= -\dot m_f. \label{eq:hyp_mdot}
\end{align}
\end{subequations}
The variables and parameters used in \eqref{eq:hyp_dyn_ct} are described below.
\noindent\textbf{Gravity model:}
Gravity is assumed to be altitude-dependent and is modeled as
\begin{equation}\label{eq:gravity}
g(h) = g_0\left(\frac{R_E}{R_E+\max\{0,h\}}\right)^2,
\end{equation}
where $g_0=9.80665$ m/s$^2$ and $R_E=6.371\times 10^6$ m.

\noindent\textbf{Atmospheric model:}
Atmospheric density $\rho(h)$, temperature $T(h)$, and speed of sound $a(h)$ are obtained
from a piecewise standard-atmosphere model \cite{USStandardAtmosphere1976}. The altitude is
clamped to $h\in[0,86{,}000]$ m. For each atmospheric layer $h\in[h_b,h_{b+1})$, the
temperature and pressure are computed as
\begin{equation}\label{eq:isa_T}
T(h)=
\begin{cases}
T_b + L_b(h-h_b), & L_b \neq 0,\\
T_b, & L_b = 0,
\end{cases}
\end{equation}
\begin{equation}\label{eq:isa_p}
p(h)=
\begin{cases}
p_b\left(\dfrac{T(h)}{T_b}\right)^{-\frac{g_0}{R_{\mathrm{air}}L_b}}, & L_b \neq 0,\\[6pt]
p_b\exp\!\left(-\dfrac{g_0(h-h_b)}{R_{\mathrm{air}}T_b}\right), & L_b = 0,
\end{cases}
\end{equation}
where $T_b$ and $p_b$ denote the temperature and pressure at the base altitude $h_b$ of the
layer, and $L_b$ is the \emph{temperature lapse rate}, i.e., the constant vertical
temperature gradient $dT/dh$ within that layer. The specific gas constant for air is
$R_{\mathrm{air}}=287.05287$~J/(kg$\cdot$K). The density and speed of sound are then
\begin{equation}\label{eq:isa_rho}
\rho(h)=\frac{p(h)}{R_{\mathrm{air}}T(h)},
\end{equation}
\begin{equation}\label{eq:isa_a}
a(h)=\sqrt{\gamma_{\mathrm{air}}R_{\mathrm{air}}T(h)},
\end{equation}
with $\gamma_{\mathrm{air}}=1.4$.

\noindent\textbf{Mach number and dynamic pressure:}
Using $a(h)$, the Mach number and dynamic pressure are defined as
\begin{equation}\label{eq:mach}
M(h,V) \triangleq \frac{V}{a(h)}.
\end{equation}
\begin{equation}\label{eq:q}
q(h,V) \triangleq \frac{1}{2}\rho(h)V^2.
\end{equation}
\begin{figure*}[h]
\centering
\subfigure[]{\includegraphics[width=0.24\linewidth]{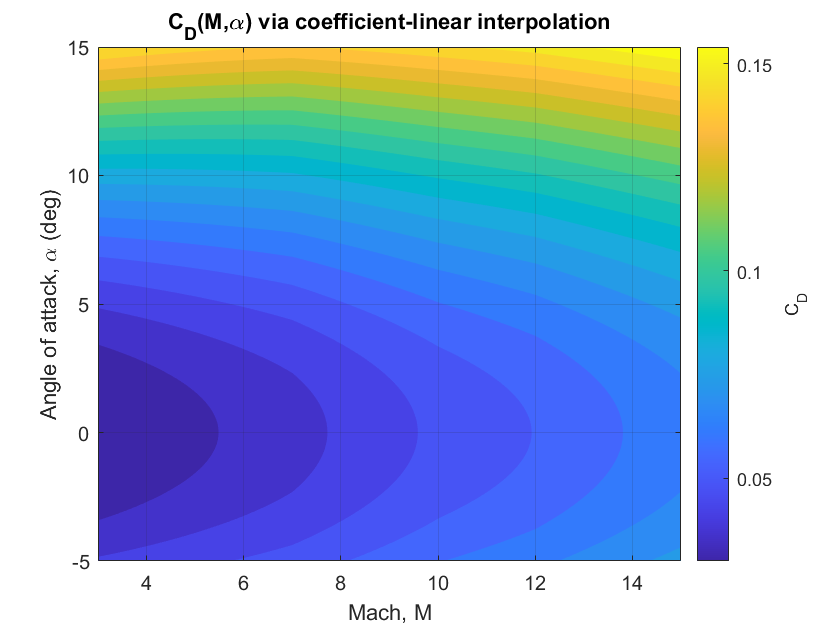}}
\subfigure[]{\includegraphics[width=0.24\linewidth]{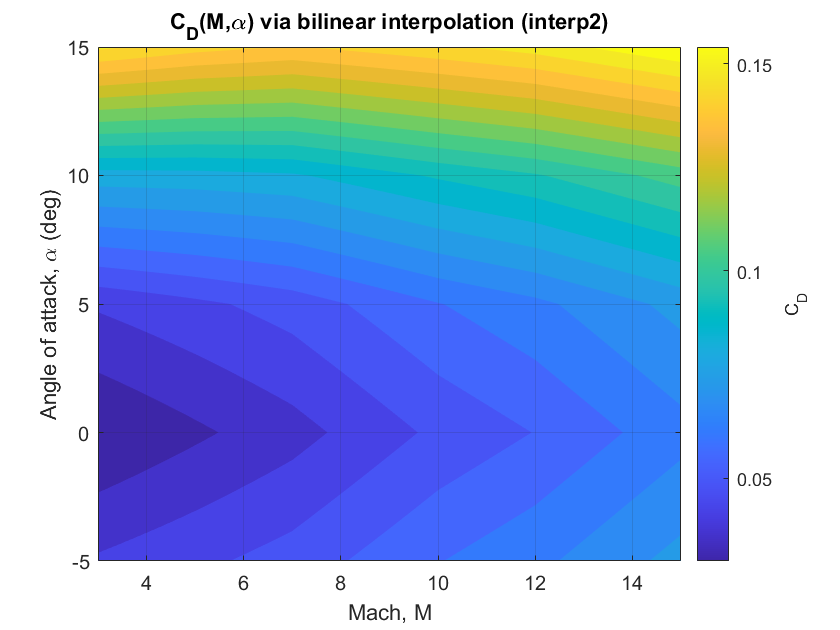}}
\subfigure[]{\includegraphics[width=0.24\linewidth]{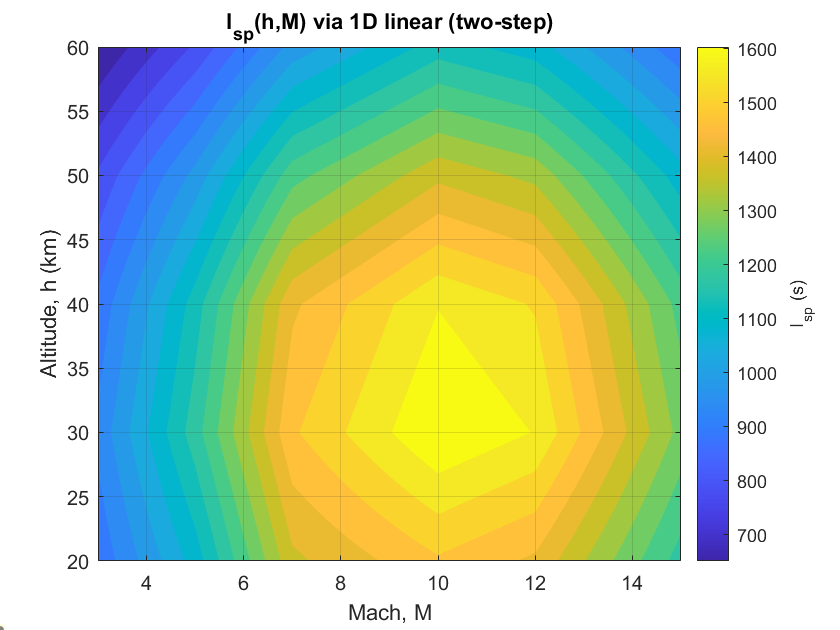}}
\subfigure[]{\includegraphics[width=0.24\linewidth]{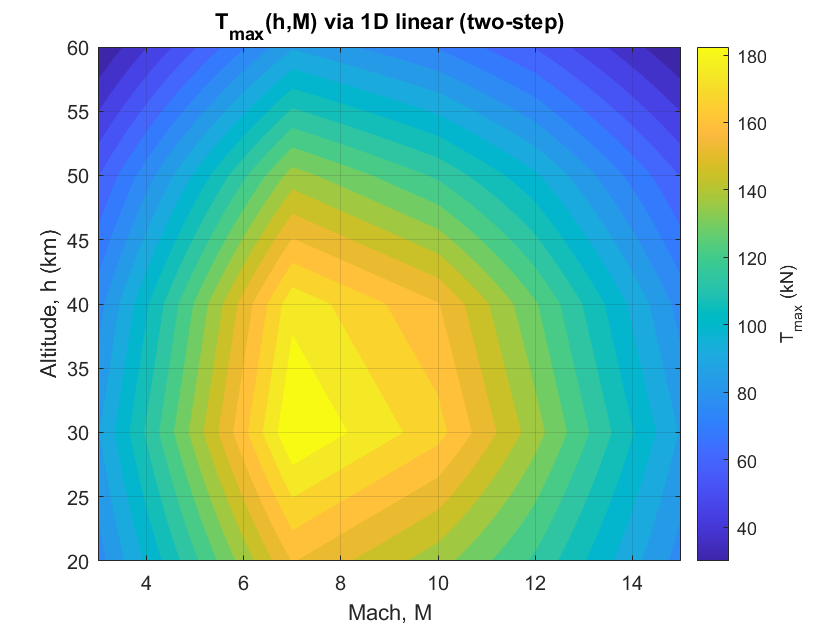}}
\caption{Contours for (a) drag coefficient, (b) lift coefficient, (c) specific impulse, and (d) maximum thrust.
}
\label{Contours}
\end{figure*}

\noindent\textbf{Aerodynamic forces:}
Lift and drag are modeled using Mach-scheduled aerodynamic coefficient maps with a quadratic drag polar \cite{Anderson2006Hypersonic,Bolender2007}.  The lift and drag coefficients are
\begin{equation}\label{eq:CL_CD}
C_L(M,\alpha) = C_{L0}(M) + C_{L\alpha}(M)\,\alpha,
\end{equation}
\begin{equation}\label{eq:CD}
C_D(M,\alpha) = C_{D0}(M) + K(M)\,C_L^2(M,\alpha) + C_{D\alpha^2}(M)\,\alpha^2.
\end{equation}

The Mach-dependent coefficients are specified on the discrete Mach grid
\[
\mathcal{M}=\{3,\,5,\,7,\,10,\,12,\,15\},
\]
which defines the tabulated Mach numbers, and are evaluated at intermediate operating
conditions via linear interpolation.
 Table~\ref{tab:aero_coeffs} summarizes the aerodynamic parameters, and Figures~\ref{Contours}(a) and~\ref{Contours}(b) illustrate the resulting \(C_L\) and \(C_D\) contours over the \(M\)–\(\alpha\) plane. The coefficients are chosen to be representative of control-oriented hypersonic vehicle models commonly used in the literature \cite{Anderson2006Hypersonic,Bolender2007,Chavez1994}. The angle of attack is constrained to \(\alpha\in[\alpha_{\min},\alpha_{\max}]\), with \(\alpha_{\min}=-5^\circ\) and \(\alpha_{\max}=15^\circ\).

The corresponding lift and drag forces are computed as
\begin{equation}\label{eq:L}
L(h,V,\alpha)=q(h,V)\,S\,C_L(M,\alpha),
\end{equation}
\begin{equation}\label{eq:D}
D(h,V,\alpha)=q(h,V)\,S\,C_D(M,\alpha).
\end{equation}
where \(S\) denotes the reference area.

\begin{table}[t]
\centering
\caption{Mach-Scheduled Aerodynamic Coefficients}
\label{tab:aero_coeffs}
\footnotesize
\renewcommand{\arraystretch}{1.1}
\begin{tabular}{ccccccc}
\hline
\textbf{Mach $M$} & $C_{L0}$ & $C_{L\alpha}$ [rad$^{-1}$] & $C_{D0}$ & $K$ & $C_{D\alpha^2}$ \\
\hline
3   & 0 & 2.8 & 0.030 & 0.120 & 0.80 \\
5   & 0 & 2.6 & 0.035 & 0.110 & 0.85 \\
7   & 0 & 2.4 & 0.040 & 0.100 & 0.90 \\
10  & 0 & 2.2 & 0.050 & 0.095 & 0.95 \\
12  & 0 & 2.1 & 0.055 & 0.090 & 1.00 \\
15  & 0 & 2.0 & 0.065 & 0.085 & 1.05 \\
\hline
\end{tabular}
\end{table}

\noindent\textbf{Aerodynamic heating proxy:}
A simplified aerodynamic heating proxy is introduced to enforce a thermal constraint,
\begin{equation}\label{eq:heat}
\dot Q(h,V) \triangleq k_{\mathrm{heat}}\sqrt{\max\{\rho(h),0\}}\;V^3,
\end{equation}
where $k_{\mathrm{heat}}>0$ is a calibrated coefficient (default $k_{\mathrm{heat}}=10^{-5}$).
\begin{table*}[t]
\centering
\caption{Maximum thrust map $T_{\max}(h,M)$ used in simulations (kN).}
\label{tab:Tmax_map}
\footnotesize
\renewcommand{\arraystretch}{1.1}
\begin{tabular}{ccccccc}
\hline
\textbf{$h$ (km)} & \textbf{$M=3$} & \textbf{$M=5$} & \textbf{$M=7$} & \textbf{$M=10$} & \textbf{$M=12$} & \textbf{$M=15$} \\
\hline
20 & 80  & 120 & 160 & 140 & 120 & 80  \\
30 & 90  & 140 & 190 & 170 & 140 & 90  \\
40 & 80  & 130 & 180 & 160 & 130 & 80  \\
50 & 60  & 100 & 140 & 120 & 100 & 60  \\
60 & 30  & 60  & 90  & 75  & 60  & 30  \\
\hline
\end{tabular}
\end{table*}
\begin{table*}[t]
\centering
\caption{Specific impulse map $I_{\mathrm{sp}}(h,M)$ used in simulations (s).}
\label{tab:Isp_map}
\footnotesize
\renewcommand{\arraystretch}{1.1}
\begin{tabular}{ccccccc}
\hline
\textbf{$h$ (km)} & \textbf{$M=3$} & \textbf{$M=5$} & \textbf{$M=7$} & \textbf{$M=10$} & \textbf{$M=12$} & \textbf{$M=15$} \\
\hline
20 & 900  & 1100 & 1400 & 1500 & 1450 & 1200 \\
30 & 950  & 1200 & 1500 & 1650 & 1600 & 1300 \\
40 & 900  & 1150 & 1450 & 1600 & 1550 & 1250 \\
50 & 800  & 1000 & 1250 & 1400 & 1350 & 1100 \\
60 & 650  & 800  & 1000 & 1150 & 1100 & 900  \\
\hline
\end{tabular}
\end{table*}

\noindent\textbf{Propulsion model and fuel flow:}
Propulsion is modeled using control-oriented, altitude- and Mach-dependent maps for the
maximum available thrust \(T_{\max}(h,M)\) and the specific impulse
\(I_{\mathrm{sp}}(h,M)\), consistent with hypersonic air-breathing propulsion trends
reported in the literature \cite{Heiser1994,Bolender2007}. The maps are defined on the grid
\[
h\in\{20,30,40,50,60\}\ \text{km}, \qquad
M\in\{3,5,7,10,12,15\},
\]
and evaluated at intermediate operating points via bilinear interpolation. More specifically, for \((h,M)\) lying between adjacent grid points \((h_i,M_j)\) and \((h_{i+1},M_{j+1})\),
both \(T_{\max}(h,M)\) and \(I_{\mathrm{sp}}(h,M)\) are computed as
\begin{equation}\label{eq:bilinear_prop}
Z(h,M)=\sum_{p\in\{i,i+1\}}\sum_{q\in\{j,j+1\}}
w_{pq}(h,M)\,Z(h_p,M_q),
\end{equation}
where \(Z\in\{T_{\max},\,I_{\mathrm{sp}}\}\) and the bilinear interpolation weights
\(w_{pq}\) satisfy \(\sum w_{pq}=1\). For \((h,M)\in[h_i,h_{i+1}]\times[M_j,M_{j+1}]\), the
weights are given by
\begin{align}
w_{i,j}     &= (1-\lambda_h)(1-\lambda_M), \nonumber\\
w_{i+1,j}   &= \lambda_h(1-\lambda_M),     \nonumber\\
w_{i,j+1}   &= (1-\lambda_h)\lambda_M,     \nonumber\\
w_{i+1,j+1} &= \lambda_h\lambda_M,
\end{align}
with
\begin{equation}
\lambda_h \triangleq \frac{h-h_i}{h_{i+1}-h_i}, \qquad
\lambda_M \triangleq \frac{M-M_j}{M_{j+1}-M_j}.
\end{equation}

The throttle input \(\delta\in[0,1]\) scales the available thrust according to
\begin{equation}\label{eq:thrust}
T(h,V,\delta)=\delta\,T_{\max}\!\big(h,M(h,V)\big).
\end{equation}
To enforce the propulsion operating envelope, the maximum thrust is set to zero outside
the nominal hypersonic Mach range,
\begin{equation}\label{eq:thrust_gate}
T_{\max}(h,M)=0 \quad \text{if } M<4 \text{ or } M>15.
\end{equation}

Fuel consumption is computed using the standard specific-impulse relation
\begin{equation}\label{eq:mdot_f}
\dot m_f=
\begin{cases}
\dfrac{T}{I_{\mathrm{sp}}(h,M)\,g_0}, & T>0,\\[4pt]
0, & T=0,
\end{cases}
\qquad \dot m=-\dot m_f,
\end{equation}
where \(g_0\) denotes standard gravity. The propulsion maps reported in
Tables~\ref{tab:Tmax_map} and~\ref{tab:Isp_map} are selected to be representative of
control-oriented hypersonic air-breathing propulsion models.

\section{Safety and Operational Constraints}\label{sec:constraints}

Let \(\mathcal{X}\subset\mathbb{R}^n\) denote a compact region of interest for the
continuous state, where \(n=4\) and
\[
x \triangleq (h,V,\gamma,m)
\]
denotes altitude, speed, flight–path angle, and mass, respectively.
Constraints are specified by a finite collection of continuously
differentiable functions
\begin{equation}
\mathcal{C}=\{c_j:\mathbb{R}^4\times\mathcal{U}\to\mathbb{R}\}_{j=1}^{N_c},
\qquad N_c=11,
\end{equation}
which are partitioned into \emph{soft} and \emph{hard} constraints as
\(\mathcal{C}=\mathcal{C}_{\mathrm{soft}}\;\dot{\cup}\;\mathcal{C}_{\mathrm{hard}}\).
The admissible control input is \(u=(\alpha,\delta)\), with control set
\begin{equation}
\mathcal{U} \triangleq [\alpha_{\min},\alpha_{\max}] \times [0,1].
\end{equation}

A constraint $c_j$ is satisfied whenever $c_j(x,u)\le 0$.
Violations of hard constraints correspond to physical failure and immediate episode
termination, whereas violations of soft constraints represent recoverable operating
conditions handled through reward shaping and/or mode-dependent logic.

\noindent\textbf{Soft constraints (operational envelopes and safety box).}
We impose $N_s=6$ soft constraints on altitude, speed, and flight--path angle,
\begin{subequations}\label{eq:soft_constraints}
\begin{align}
c_{1}(x) &= h_{\min}-h, & c_{2}(x) &= h-h_{\max}, \\
c_{3}(x) &= V_{\min}-V, & c_{4}(x) &= V-V_{\max}, \\
c_{5}(x) &= \gamma_{\min}-\gamma, &
c_{6}(x) &= \gamma-\gamma_{\max}.
\end{align}
\end{subequations}
These constraints define operational envelopes and do not represent physical failure.
In the implemented environment, they are not used to terminate episodes nor to
define hard feasibility.

Instead, a compact \emph{safety box}
$\mathcal{B}_{\mathrm{safe}}\subset(h,V,\gamma)$ centered at a nominal operating point
$(h_{\mathrm{nom}},V_{\mathrm{nom}},\gamma_{\mathrm{nom}})$ is defined as
\begin{equation}\label{eq:safetybox}
\resizebox{0.99\hsize}{!}{%
$
\mathcal{B}_{\mathrm{safe}}
=
\bigl\{
x:\;
|h-h^*|\le \Delta h,\;
|V-V^*|\le \Delta V,\;
|\gamma-\gamma^*|\le \Delta\gamma
\bigr\},
$
}
\end{equation}
where $(h^*,V^*,\gamma^*)$ defines the reference altitude, speed, and flight-path angle. 
Membership in $\mathcal{B}_{\mathrm{safe}}$ is used for mode-dependent reward shaping
and analysis, but leaving the box is permitted and does not constitute failure.

\noindent\textbf{Hard constraints (physical failure conditions).}
The normal load factor induced by the commanded angle of attack is
\[
n(h,V,\alpha,m)\triangleq \frac{L(h,V,\alpha)}{m\,g(h)}.
\]
The hard constraints are
\begin{subequations}\label{eq:hard_constraints}
\begin{align}
c_{7}(x)   &= q(h,V)-q_{\max}, \\
c_{8}(x,u) &= |n(h,V,\alpha,m)|-n_{\max}, \\
c_{9}(x)   &= \dot Q(h,V)-\dot Q_{\max},\\
c_{10}(x)  &= M(h,V)-M_{\max},\\
c_{11}(x)  &= M_{\min}-M(h,V),
\end{align}
\end{subequations}
with $M_{\min}=4$ and $M_{\max}=15$.
Violating any constraint $c_j\in\mathcal{C}_{\mathrm{hard}}$ results in immediate
episode termination.

\begin{definition}[One-step hard-safe action]\label{def:hardsafe}
Let $f_\Delta(x,u)$ denote the RK2-discretized dynamics with time step $\Delta t$.
An action $u\in\mathcal{A}$ is \emph{hard-safe} at state $x\in\mathcal{X}$ if
\[
c_j\!\left(f_\Delta(x,u),u\right)\le 0,
\qquad \forall\, c_j\in\mathcal{C}_{\mathrm{hard}}.
\]
\end{definition}

\begin{definition}[State aggregation and representative states]\label{def:aggregation}
Let $\Pi:\mathcal{X}\to\mathcal{S}$ denote the state aggregation (binning) operator used
to construct a finite abstract state space.
Each abstract state $s\in\mathcal{S}$ is associated with a representative continuous
state $x_{\mathrm{rep}}(s)$ at which one-step transitions are evaluated.
\end{definition}

\begin{definition}[Viable feasible set]\label{def:feas}
Define \(\mathcal{S}_{\mathrm{feas}}\subseteq\mathcal{S}\) as the largest subset such that
for every \(s\in\mathcal{S}_{\mathrm{feas}}\) there exists at least one action
\(u\in\mathcal{A}\) satisfying
\[
c_j\!\left(f_\Delta(x_{\mathrm{rep}}(s),u),u\right)\le 0,
\quad \forall c_j\in\mathcal{C}_{\mathrm{hard}},
\]
and
\[
\Pi\!\left(f_\Delta(x_{\mathrm{rep}}(s),u)\right)\in\mathcal{S}_{\mathrm{feas}}.
\]
Equivalently, \(\mathcal{S}_{\mathrm{feas}}\) is obtained as the fixed point of an
iterative pruning procedure that removes any abstract state for which no such
hard-safe action exists.
\end{definition}

\begin{definition}[Admissible action set]\label{def:Amask}
Let \(\mathcal{A}\subset\mathcal{U}\) denote a finite set of control inputs (actions)
that can be executed by the vehicle. For each
\(s\in\mathcal{S}_{\mathrm{feas}}\), the admissible action set
\(\mathcal{A}_{\mathrm{mask}}(s)\subseteq\mathcal{A}\) is defined as
\begin{equation}\label{eq:Amask}
\begin{split}
\mathcal{A}_{\mathrm{mask}}(s)
\triangleq
\Bigl\{
u\in\mathcal{A}:\;
c_j\!\left(f_\Delta(x_{\mathrm{rep}}(s),u),u\right)\le 0,\ \\
\forall\, c_j\in\mathcal{C}_{\mathrm{hard}},\;
\Pi\!\left(f_\Delta(x_{\mathrm{rep}}(s),u)\right)\in\mathcal{S}_{\mathrm{feas}}
\Bigr\}.
\end{split}
\end{equation}
This viability-based admissible action set eliminates hard-safe dead-end conditions
by construction and is used consistently for both action selection and value updates.
\end{definition}

\noindent\textbf{Soft-safe classification.}
For analysis and reward-mode selection, a feasible abstract state $s$ may be labeled
\emph{soft-safe} if its representative state satisfies
$c_j(x_{\mathrm{rep}}(s))\le 0$ for all $c_j\in\mathcal{C}_{\mathrm{soft}}$,
and \emph{soft-unsafe} otherwise.
Feasibility depends solely on hard constraints and is independent of this label.

\section{Hybrid Control Framework}\label{HybridFramework}

For each constraint \(c_j\in\mathcal{C}\), define the violation indicator
\begin{equation}
v_j(x,u)
\triangleq
\mathbf{1}\!\left\{c_j(x,u)>0\right\},
\qquad j=1,\ldots,N_c,
\end{equation}
and collect them into the violation vector
\begin{equation}
v(x,u)\triangleq (v_1(x,u),\ldots,v_{N_c}(x,u))\in\{0,1\}^{N_c}.
\end{equation}

Violations of hard constraints correspond to physical failure and immediate episode
termination, whereas violations of soft constraints indicate recoverable operating
conditions and are handled through mode-dependent reward shaping.

Let $\Pi:\mathcal{X}\to\mathcal{S}$ denote a state aggregation (projection) operator
that maps a continuous state $x$ to a discrete abstract state $s=\Pi(x)$.
Based on the reachability analysis described in Section~\ref{sec:constraints},
the effective abstract state space is restricted to the finite set
$\mathcal{S}_{\mathrm{feas}}\subset\mathcal{S}$ of \emph{feasible} states, i.e.,
states that admit at least one hard-safe action whose successor remains feasible.
Abstract states corresponding to hard-safe dead-end conditions are excluded from
$\mathcal{S}_{\mathrm{feas}}$ by construction.

\noindent\textbf{Mode definition and state partition.}
The feasible abstract state space $\mathcal{S}_{\mathrm{feas}}$ is partitioned into two
disjoint subsets according to membership in the safety box:
\begin{align}
\mathcal{S}_{\mathrm{safe}}
&:= \bigl\{\, s\in\mathcal{S}_{\mathrm{feas}} \;\big|\;
x_{\mathrm{rep}}(s)\in\mathcal{B}_{\mathrm{safe}} \,\bigr\}, \\
\mathcal{S}_{\mathrm{unsafe}}
&:= \mathcal{S}_{\mathrm{feas}}\setminus \mathcal{S}_{\mathrm{safe}} ,
\end{align}
where $x_{\mathrm{rep}}(s)$ denotes the representative continuous state associated with
$s$.
The discrete operating mode at time step $k$ is defined as
\[
\sigma_k =
\begin{cases}
\mathrm{safe}, & s_k \in \mathcal{S}_{\mathrm{safe}},\\
\mathrm{unsafe}, & s_k \in \mathcal{S}_{\mathrm{unsafe}}.
\end{cases}
\]

\noindent\textbf{Augmented-state MDP interpretation.}
For analysis, the closed-loop system is viewed as a Markov decision process over the
augmented state
\[
\tilde{s}_k := (s_k,\sigma_k) \in \mathcal{S}_{\mathrm{feas}} \times \Gamma,
\qquad
\Gamma=\{\mathrm{safe},\mathrm{unsafe}\}.
\]
The resulting MDP is
\[
\mathcal{M}
=
\bigl(
\mathcal{S}_{\mathrm{feas}}\times\Gamma,\;
\mathcal{A},\;
\mathcal{A}_{\mathrm{mask}},\;
\mathcal{P},\;
\mathcal{R},\;
\gamma
\bigr),
\]
where $\mathcal{A}\subset \mathcal{U}$ is the discrete action set and $\gamma\in[0,1]$ is the discount
factor.
In the implementation, learning and action selection depend only on the abstract
state $s_k$, while $\sigma_k$ is used to select the appropriate reward expression.

\noindent\textbf{Admissible action set.}
The admissible action mapping
\(
\mathcal{A}_{\mathrm{mask}}:\mathcal{S}_{\mathrm{feas}}\to 2^{\mathcal{A}}
\)
enforces hard safety and feasibility constraints.
Specifically, for any $s\in\mathcal{S}_{\mathrm{feas}}$,
\begin{itemize}
\item all admissible actions are hard-safe in the one-step sense; and
\item all admissible actions have one-step successors that remain in
$\mathcal{S}_{\mathrm{feas}}$.
\end{itemize}
The admissible action set does not depend explicitly on the mode $\sigma_k$; recovery
from unsafe states and regulation within the safety box are instead driven by the
mode-dependent reward.

\noindent\textbf{Transitions and rewards.}
The transition kernel
\[
\mathcal{P}:\mathcal{S}_{\mathrm{feas}}\times\mathcal{A}_{\mathrm{mask}}
\;\to\;
\mathcal{S}_{\mathrm{feas}}
\]
is induced by the discretized vehicle dynamics and the projection operator $\Pi$.
The reward function
\(
\mathcal{R}:\mathcal{S}_{\mathrm{feas}}\times\mathcal{A}\times
\mathcal{S}_{\mathrm{feas}} \to \mathbb{R}
\)
is explicitly \emph{mode-dependent}:
distinct reward expressions are used depending on whether the current state lies in
$\mathcal{S}_{\mathrm{safe}}$ or $\mathcal{S}_{\mathrm{unsafe}}$.

\subsection{Dynamics Discretization, State Abstraction, and Reward Design}
\label{DynamicsAbstraction}

Given a sampling period $\Delta t>0$, the continuous-time dynamics $\dot{x}=g(x,u)$ are discretized
using an explicit second-order Runge--Kutta (midpoint) scheme:
\begin{align}
k_1 &= g(x_k,u_k), \nonumber\\
x_{k+\frac12} &= x_k + \tfrac{\Delta t}{2}k_1, \nonumber\\
k_2 &= g(x_{k+\frac12},u_k), \nonumber\\
x_{k+1} &= x_k + \Delta t\,k_2. \label{eq:rk2}
\end{align}

To obtain a finite Markov decision process (MDP), the continuous state is mapped to a discrete abstract
state via the projection $\Pi:\mathcal{X}\to\mathcal{S}$. 
The nominal action space $\mathcal{A}$ is identical across modes.

\noindent\textbf{Safe mode (inside the safety box).}
When $\sigma_k=\mathrm{safe}$, the reward promotes regulation about the nominal
operating condition $(h^*,V^*,\gamma^*)$ while penalizing control effort and excessive
command switching. The reward is defined as
\begin{equation}\label{reward-safe}
\mathcal{R}_{\mathrm{safe}}(x_k,a_k)
=
- \ell_{\mathrm{trk}}(x_{k+1})
- w_u \lVert a_k\rVert^2
- \ell_{\mathrm{sw}}(a_k,a_{k-1}),
\end{equation}
where the tracking loss is given by
\begin{equation}
\ell_{\mathrm{trk}}(x)=
w_h (h - h^*)^2
+ w_V (V - V^*)^2
+ w_\gamma (\gamma - \gamma^*)^2 ,
\end{equation}
and $w_h$, $w_V$, and $w_\gamma$ are positive weighting coefficients that balance
tracking performance in altitude, speed, and flight--path angle, respectively, and
$w_u>0$ penalizes control effort.

To discourage rapid changes in the control input near the nominal operating condition,
a switching penalty
\begin{equation}\label{lsw}
\ell_{\mathrm{sw}}(a_k,a_{k-1})
=
w_{du}\!\left(
(\alpha_k-\alpha_{k-1})^2
+
\lambda_\delta(\delta_k-\delta_{k-1})^2
\right)
\end{equation}
is included, where $w_{du}>0$ and $\lambda_\delta>0$ weight changes in angle of attack
and throttle, respectively. This term may be disabled by setting $w_{du}=0$.

\noindent\textbf{Unsafe mode (outside the safety box).}
When $\sigma_k=\mathrm{unsafe}$, the reward encourages recovery toward the safety box
while continuing to penalize control effort:
\begin{equation}\label{reward-unsafe}
\resizebox{0.99\hsize}{!}{%
$
\begin{split}
\mathcal{R}_{\mathrm{unsafe}}(x_k,a_k)
=&
- c_{\mathrm{out}}
- w_d\, d(x_k)
- w_u \lVert a_k\rVert^2 - \ell_{\mathrm{sw}}(a_k,a_{k-1})\\
- &c_{\mathrm{away}}\,\mathbf{1}\!\left\{d(x_{k+1})>d(x_k)\right\}
 +w_{\mathrm{imp}}\!\left[d(x_k)-d(x_{k+1})\right]_+,
\end{split}
$
}
\end{equation}
where $c_{\mathrm{out}}>0$ is a constant offset penalty applied outside the safety box,
$[\cdot]_+=\max\{\cdot,0\}$ denotes the positive-part operator, and
$\mathbf{1}\{\cdot\}$ is the indicator function. The distance $d(x)$ measures the
normalized deviation from the safety-box center,
\begin{equation}\label{eq:dist}
d(x)=
\sqrt{
\left(\frac{h-h^*}{\Delta h}\right)^2
+\left(\frac{V-V^*}{\Delta V}\right)^2
+\left(\frac{\gamma-\gamma^*}{\Delta\gamma}\right)^2 }.
\end{equation}
The coefficients $w_d>0$, $w_{\mathrm{imp}}>0$, and $c_{\mathrm{away}}>0$ respectively
penalize large deviations from the nominal operating condition, reward progress toward
the safety box, and discourage actions that increase the distance from its center.

\subsection{RL with Shielding and Episode Chaining}
\label{RLShieldChain}

We employ tabular Q-learning to compute a stationary policy
\begin{equation}
\pi:\mathcal{S}_{\mathrm{feas}}\times\Gamma\rightarrow\mathcal{A},
\end{equation}
defined over the finite augmented abstract state space
\(\mathcal{S}_{\mathrm{feas}}\times\Gamma\).

\noindent\textbf{Shielded action space.}
At each time step $k$, the agent is restricted to the admissible action set
$\mathcal{A}_{\mathrm{mask}}(s_k)$, consisting of actions that are
one-step hard-safe and whose successors remain in the feasible abstract set.
All exploration and exploitation are confined to this set, ensuring that
hard-constraint violations are never intentionally selected.

\noindent\textbf{Masked and local action selection.}
Action selection follows a masked $\epsilon$-greedy strategy, further refined
by a \emph{local neighborhood search} to reduce large inter-step command changes
and reflect incremental actuator behavior.

Let $a_{k-1}\in\mathcal{A}$ denote the previously applied discrete action.
The discrete action space is endowed with a graph structure induced by adjacency
in the $(\alpha,\delta)$ grid, where edges connect actions differing by one grid
step in either component.
Let $\mathcal{N}^{(r)}(a_{k-1})$ denote the $r$-hop neighborhood of $a_{k-1}$ on
this action graph, with $\mathcal{N}^{(1)}(a_{k-1})$ representing immediate neighbors.

At state $s_k$, we define the smallest neighborhood radius
\begin{equation}
r_k
:=
\min\Bigl\{
r\ge 1 \;\big|\;
\mathcal{A}_{\mathrm{mask}}(s_k)\cap\mathcal{N}^{(r)}(a_{k-1})\neq\emptyset
\Bigr\},
\end{equation}
up to a prescribed maximum radius.
The \emph{local admissible action set} is then defined as
\begin{equation}
\mathcal{A}_{\mathrm{loc}}(s_k,a_{k-1})
:=
\mathcal{A}_{\mathrm{mask}}(s_k)\cap\mathcal{N}^{(r_k)}(a_{k-1}).
\end{equation}
If no admissible action is found within the maximum neighborhood radius,
the selection falls back to the full admissible set $\mathcal{A}_{\mathrm{mask}}(s_k)$.

With probability $\epsilon$, exploration samples uniformly from
$\mathcal{A}_{\mathrm{loc}}(s_k,a_{k-1})$.
With probability $1-\epsilon$, exploitation selects
\[
a_k \in
\arg\max_{a\in\mathcal{A}_{\mathrm{loc}}(s_k,a_{k-1})} Q(s_k,a).
\]
This local-first selection strategy preserves hard safety by construction
while discouraging large step-to-step changes in the discrete control command,
thereby reducing chattering and improving closed-loop smoothness.

\noindent\textbf{Mask-consistent value updates.}
The action-value function $Q:\mathcal{S}\times\mathcal{A}\to\mathbb{R}$ is updated
using a mask-consistent Bellman backup:
\begin{equation}
\resizebox{0.99\hsize}{!}{%
$
Q(s_k,a_k)\leftarrow Q(s_k,a_k)
+\alpha\Bigl(
r_k
+\gamma
\max\limits_{a\in\mathcal{A}_{\mathrm{mask}}(s_{k+1})}
Q(s_{k+1},a)
-
Q(s_k,a_k)
\Bigr).
$
}
\end{equation}
The maximization at the successor state is explicitly restricted to the
admissible action set $\mathcal{A}_{\mathrm{mask}}(s_{k+1})$.
This restriction prevents optimistic value propagation through infeasible actions
and ensures consistency between action selection and value updates. Importantly, the local action set $\mathcal{A}_{\mathrm{loc}}$ is used \emph{only}
for action selection.
Value propagation always considers the full admissible set
$\mathcal{A}_{\mathrm{mask}}$, ensuring that the learned value function reflects
the true feasible decision space.

\noindent\textbf{Episode horizon.}
Each episode is executed for a fixed horizon of $N_{\max}=400$ steps, corresponding
to a total simulated duration of $N_{\max}\Delta t$.
Episodes may also terminate earlier due to operational termination guards or
hard-constraint violations.

\noindent\textbf{Episode chaining.}
To improve data efficiency and enable long-horizon recovery behavior, we employ
an \emph{episode chaining} mechanism.
When an episode terminates at time $T$ due to reaching the horizon or non-failure
termination conditions, the initial state of the next episode is set to the terminal
continuous state of the previous episode:
\[
x_0^{(k+1)} = x_{N_{\max}}^{(k)}.
\]
If termination occurs due to violation of a hard physical constraint,
the environment is instead reset to a fixed nominal safe state
$x_{\mathrm{safe}}^{\mathrm{nom}}\in\mathcal{B}_{\mathrm{safe}}$.
Formally,
\begin{equation}
\resizebox{0.99\hsize}{!}{%
$
x_0^{(k+1)} :=
\begin{cases}
x_{N_{\max}}^{(k)}, & \text{if termination is not due to hard constraint violation},\\
x_{\mathrm{safe}}^{\mathrm{nom}}, & \text{otherwise}.
\end{cases}
$
}
\end{equation}

Episode chaining preserves the Markov property of the learning process, since each
reset depends only on the terminal state of the preceding episode.
This mechanism allows the agent to experience extended recovery and regulation
dynamics across episode boundaries while maintaining a well-defined MDP structure.

\begin{table}[t]
\centering
\caption{Discrete action values used in simulation}
\label{tab:actions}
\begin{tabular}{c c}
\hline
Angle of attack $\alpha$ (deg) & Throttle $\delta$ \\ \hline
$3,\;5,\;8,\;12,\;15$ & $0.25,\;0.50,\;0.75,\;1.00$ \\
\hline
\end{tabular}
\end{table}

\section{Simulation Environment and Conditions}
\label{sec:simulation}

In simulation, the continuous longitudinal state
$x=(h,V,\gamma,m)$ is discretized to
form a finite abstract state space. The resulting abstract state space contains on the order of $10^4$ states prior to
pruning.
An offline reachability analysis followed by fixed-point pruning removes abstract
states that cannot reach the designated goal set through hard-safe actions, yielding a
strict subset $\mathcal{S}_{\mathrm{feas}}\subset\mathcal{S}$ on which learning is
performed.

The discrete action space is defined as a Cartesian grid over angle of attack and
throttle,
\(
\mathcal{A}=\mathcal{A}_\alpha\times\mathcal{A}_\delta,
\)
with $|\mathcal{A}_\alpha|=5$ angle-of-attack levels and $|\mathcal{A}_\delta|=4$
throttle levels, for a total of $|\mathcal{A}|=20$ discrete actions (see Table \ref{tab:actions}).
For each feasible abstract state $s$, the admissible action set
$\mathcal{A}_{\mathrm{mask}}(s)$ is obtained by excluding actions that either violate
hard physical constraints in one step or transition to an infeasible abstract state.
The cardinality of $\mathcal{A}_{\mathrm{mask}}(s)$ is state dependent and satisfies
$1\le |\mathcal{A}_{\mathrm{mask}}(s)|\le 20$, with fewer admissible actions near
constraint boundaries and a larger set retained in the interior of the feasible region.
This masked action set is enforced uniformly during exploration, exploitation, and
value updates.

The nominal hypersonic cruise condition used in the simulations is
\[
(h^*,\,V^*,\,\gamma^*)
=
(35{,}000~\mathrm{m},\; 2{,}500~\mathrm{m/s},\; 0~\mathrm{rad}),
\]
with initial mass $m_0 = 12{,}000~\mathrm{kg}$.
A safety envelope over $(h,V,\gamma)$ is defined by~\eqref{eq:safetybox} and centered
at the nominal operating condition, with half-widths
\[
\Delta h = 8{,}000~\mathrm{m}, ~
\Delta V = 800~\mathrm{m/s}, ~
\Delta \gamma = 5^\circ \; (\approx 0.087~\mathrm{rad}).
\]

To enable long-horizon learning and recovery behavior, episode chaining is employed:
when an episode terminates without violating hard physical constraints, the terminal
continuous state is used as the initial condition for the subsequent episode; if a
hard constraint violation occurs, the system is reinitialized at the nominal operating
condition.
\subsection{Reward Structure}
\label{subsec:reward}
For both $\sigma=\mathrm{safe}$ and $\sigma=\mathrm{unsafe}$, the switching penalty, defined by Eq.~\eqref{lsw}, is included to mitigate action chattering with $w_{du}=5\times10^{-3}$ and $\lambda_\delta=25$. When $\sigma_k=\mathrm{safe}$, the reward promotes regulation about the nominal cruise condition $(h^,V^,\gamma^*)$ while penalizing control effort and excessive command switching. The reward is given by \eqref{reward-safe}, where the tracking loss is evaluated at the successor state $x_{k+1}$. The weighting coefficients are $w_h = 4.0\times10^{-8}$, $w_V = 4.0\times10^{-6}$, $w_\gamma = 0.04$, and $w_u = 10^{-4}$. When $\sigma_k=\mathrm{unsafe}$, the reward encourages recovery toward the safety box while continuing to penalize control effort. The reward is given by \eqref{reward-unsafe} with parameters $c_{\mathrm{out}} = 50$, $w_d = 25$, $w_{\mathrm{imp}} = 150$, and $c_{\mathrm{away}} = 10$.

\begin{figure*}[t]
    \centering
    \includegraphics[width=0.32\textwidth]{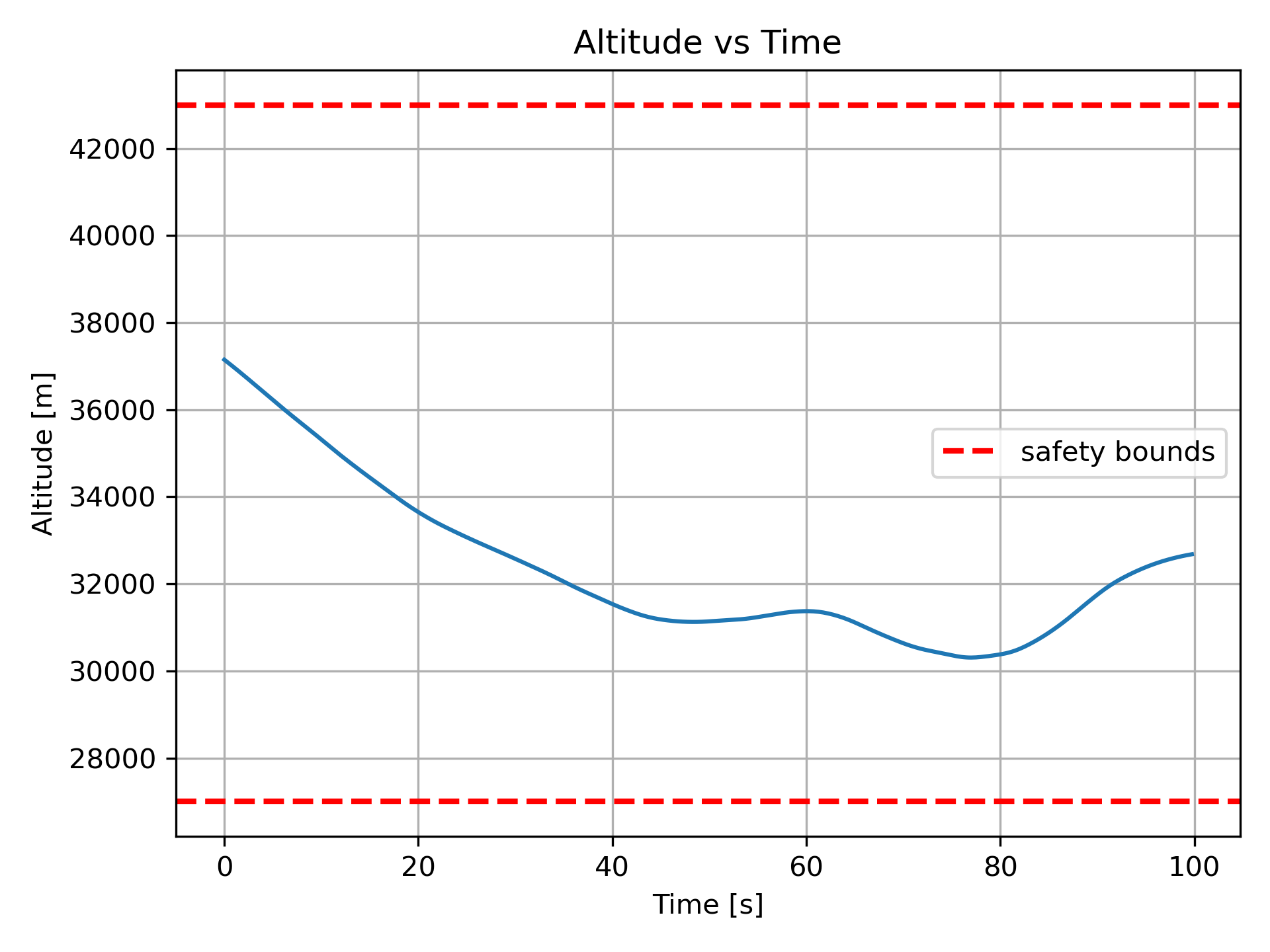}
    \includegraphics[width=0.32\textwidth]{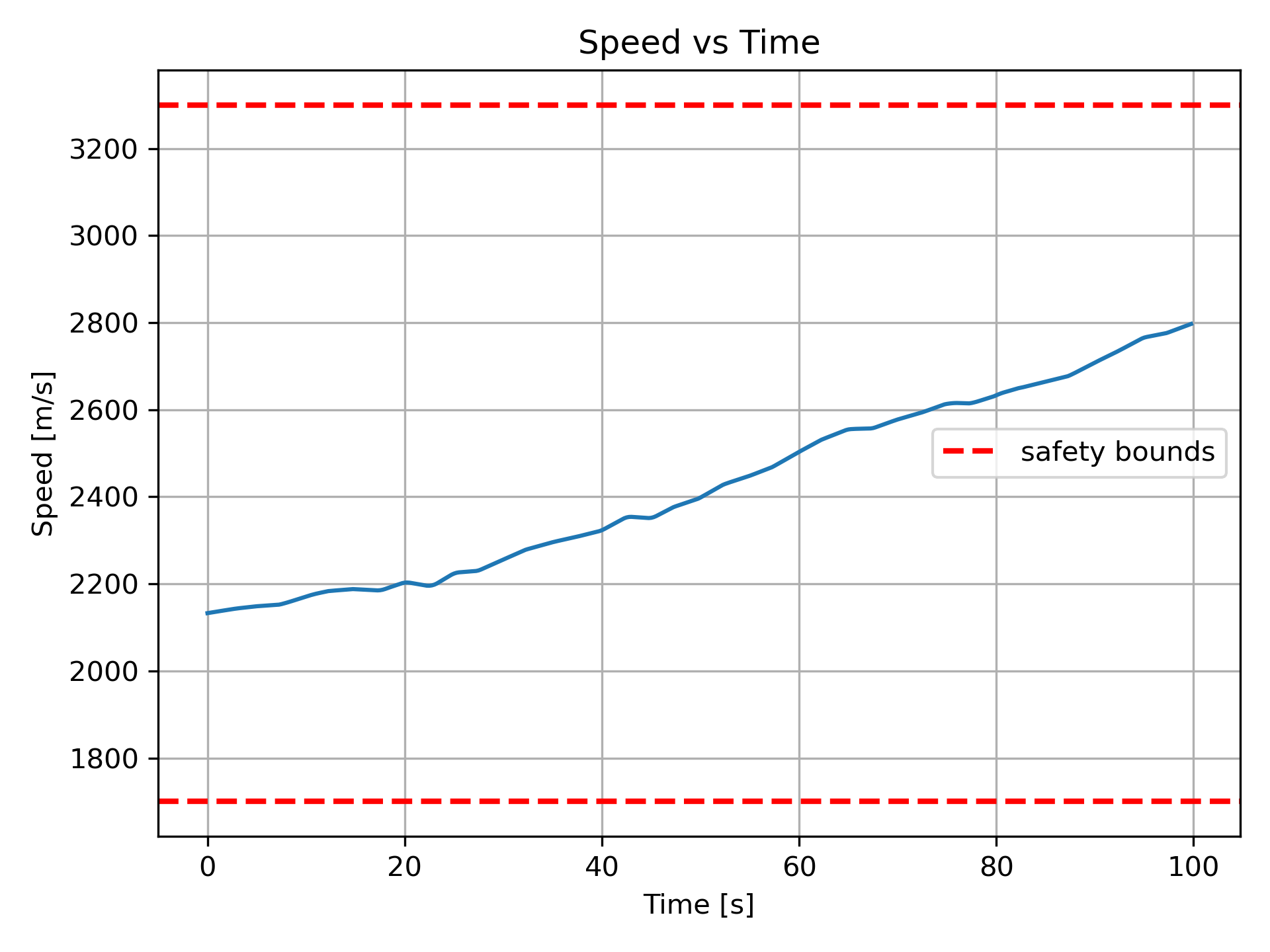}
    \includegraphics[width=0.32\textwidth]{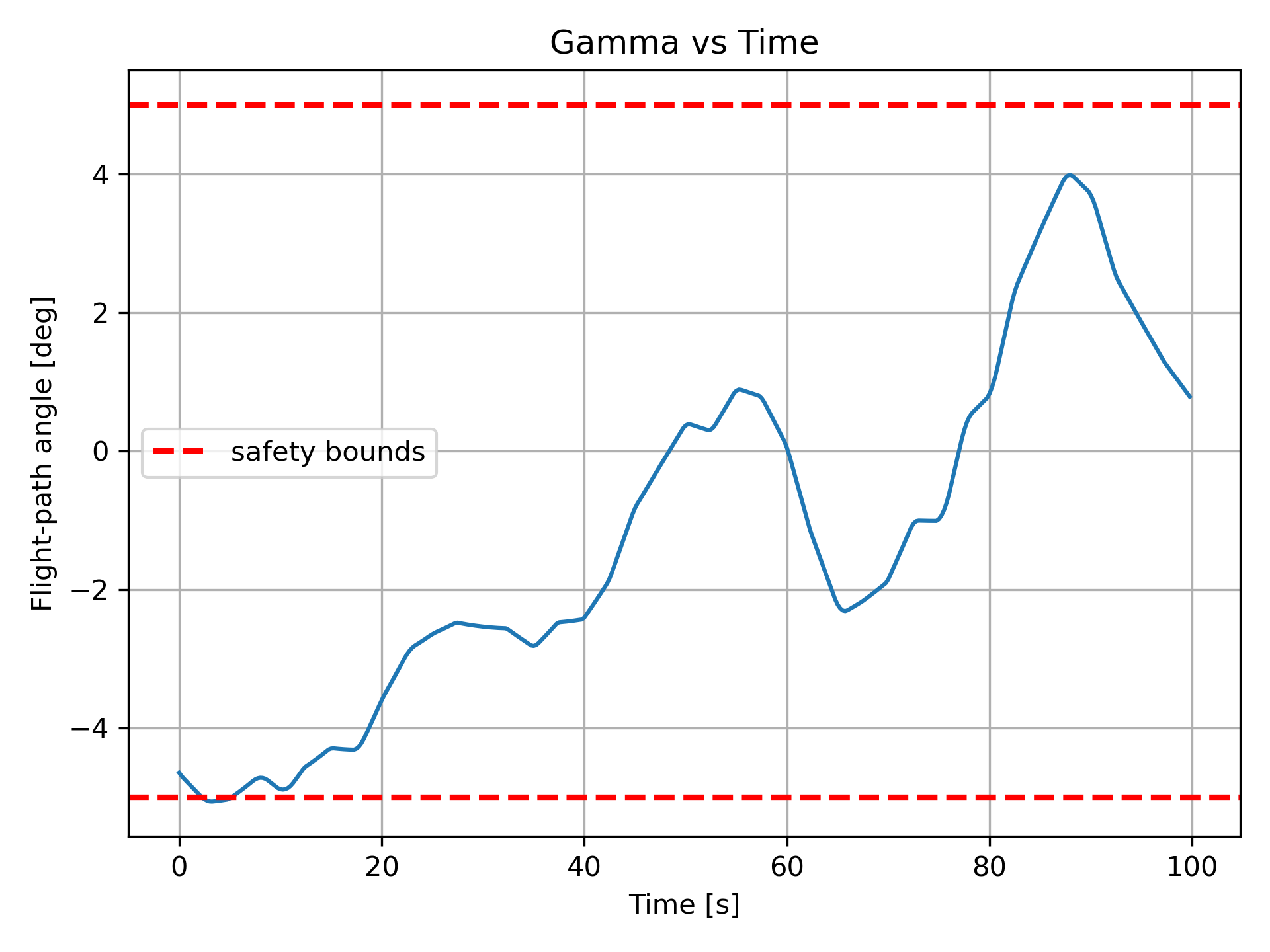}
    \caption{State trajectories starting from an initial condition outside the safety box.
    Altitude $h$, speed $V$, and flight--path angle $\gamma$ are shown.
    Red dashed lines indicate the safety--box bounds.
    The trajectory is driven back into the safe region and converges toward the nominal
    cruise condition while remaining within hard physical constraints.}
    \label{fig:state_recovery}
\end{figure*}
\subsection{Plots}
Figure~\ref{fig:state_recovery} shows the evolution of the longitudinal state starting from
an initial condition outside the safety box.
The red dashed lines indicate the safety--box bounds used for mode classification.
At the initial time, the flight--path angle $\gamma$ violates the safety envelope,
while altitude and speed remain within acceptable ranges.

Under the learned policy with reachability--based action masking, the system exhibits
a recovery maneuver that drives the state back into the safety box.
The flight--path angle is regulated toward zero, followed by gradual convergence of
altitude and speed toward their nominal cruise values.
Once the state enters the safety box, it remains inside the safe region for the remainder
of the rollout, demonstrating forward invariance induced by the admissible action set.
No violations of hard aerothermal, Mach, or load constraints are observed.

\begin{figure*}[t]
    \centering
    \includegraphics[width=0.32\textwidth]{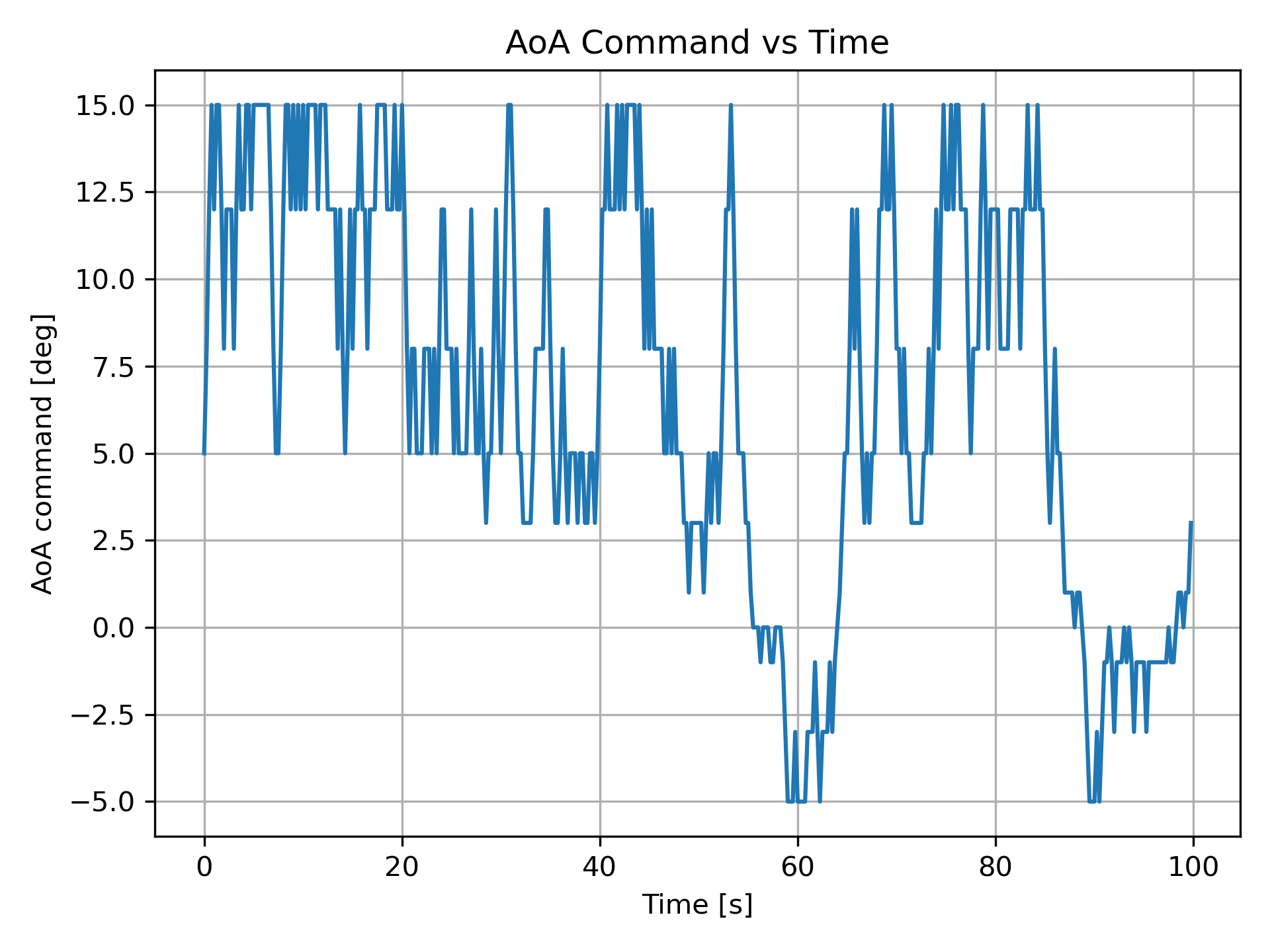}
    \includegraphics[width=0.32\textwidth]{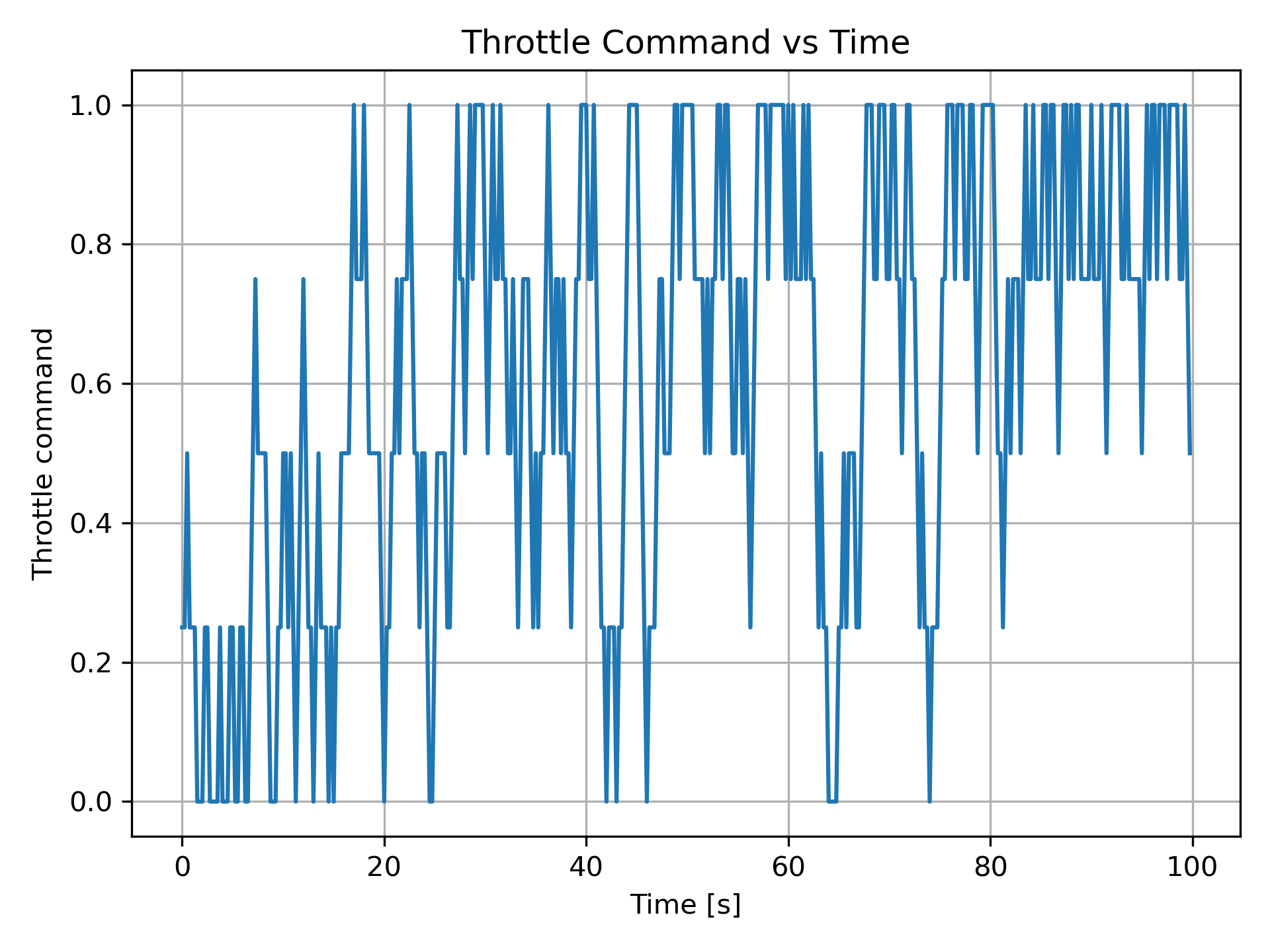}
    \includegraphics[width=0.32\textwidth]{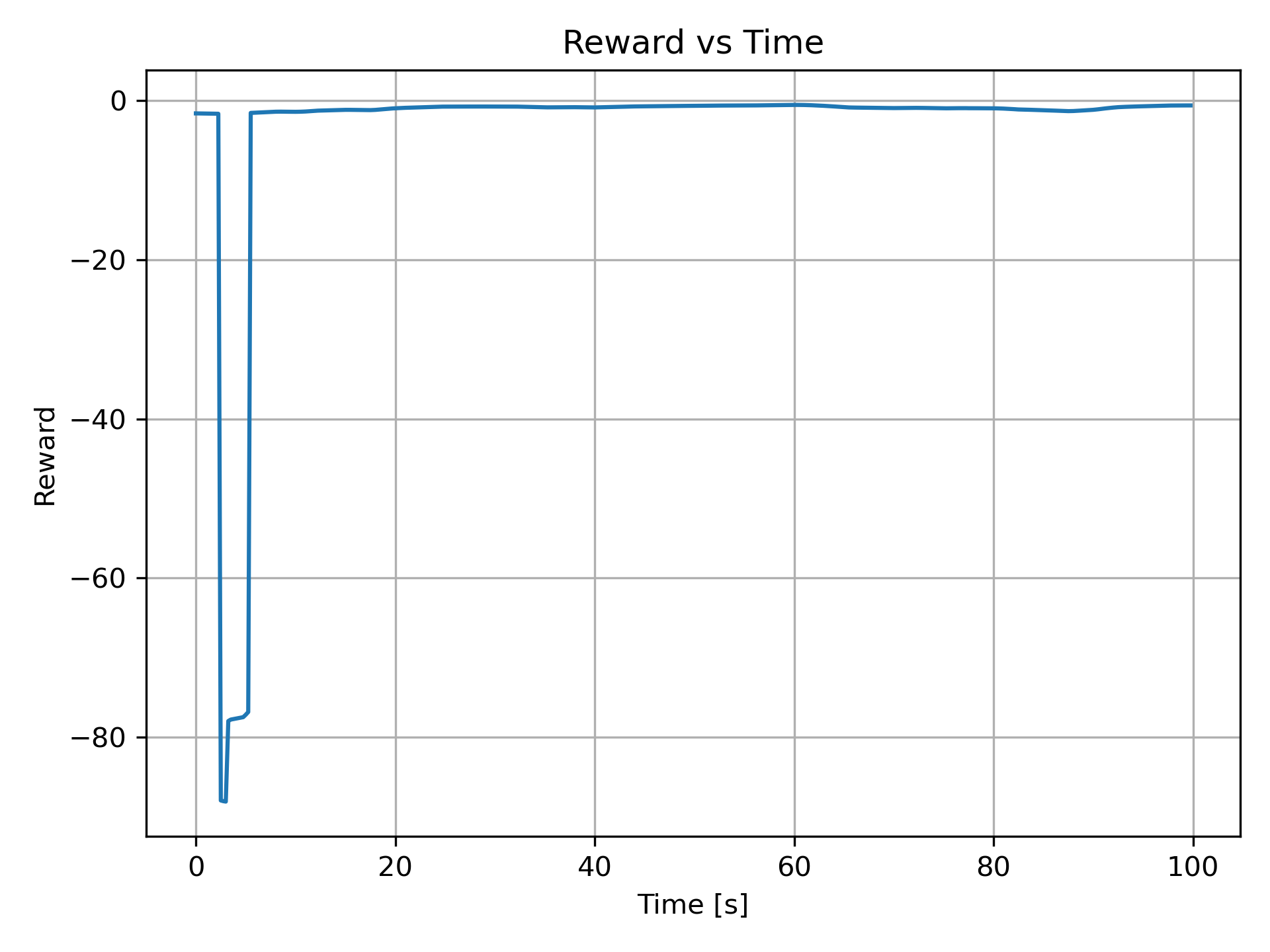}
    \caption{Control inputs and reward evolution.
    Angle--of--attack and throttle commands remain within the admissible discrete action set.
    The reward increases as the state approaches the safety box and stabilizes during
    nominal cruise operation.}
    \label{fig:actions_reward}
\end{figure*}
Fig.~\ref{fig:actions_reward} shows the commanded angle of attack, throttle, and the
instantaneous reward along the same rollout.
The control inputs respect the discrete action limits imposed by the admissible action
set and exhibit switching consistent with neighborhood--constrained action selection.
No inadmissible actions are executed. During the initial recovery phase, the reward assumes large negative values due to
penalties associated with being outside the safety box.
As the state approaches the nominal operating condition, the reward improves and
stabilizes near zero, indicating successful transition to steady cruise.

\section{Conclusion and Future Work}\label{Conclusion}

This paper presented a safety-critical reinforcement learning framework for hypersonic longitudinal flight control that integrates online tabular learning with explicit enforcement of hard safety constraints. By formulating flight control as a single constrained Markov decision process, safety is guaranteed structurally through reachability-based admissible action sets and action shielding, ensuring that both learning and execution remain within physically viable regions of the state space at all times. A finite abstraction of the continuous dynamics enables online tabular Q-learning, while mask-consistent value updates and neighborhood-based local action selection maintain compatibility between learning, safety enforcement, and actuator smoothness. Episode chaining further enhances data efficiency and allows the agent to experience long-horizon recovery behaviors without violating feasibility. Future work will extend the framework to higher-dimensional vehicle models, incorporate uncertainty and robustness considerations, and investigate function-approximation methods that preserve the formal safety guarantees provided by shielding and admissible action masking.

\bibliographystyle{IEEEtran}
\bibliography{reference}

\begin{IEEEbiography}[{\includegraphics[width=1in,height=1.25in,clip,keepaspectratio]{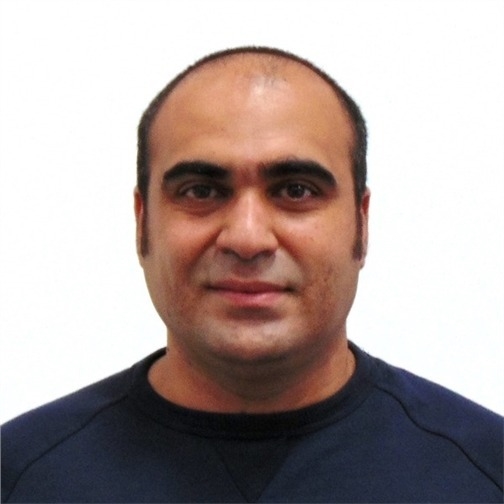}}]
{\textbf{Hossein Rastgoftar}} an Assistant Professor at the University of Arizona. Prior to this, he was an adjunct Assistant Professor at the University of Michigan from 2020 to 2021. He was also an Assistant Research Scientist (2017 to 2020) and a Postdoctoral Researcher (2015 to 2017) in the Aerospace Engineering Department at the University of Michigan Ann Arbor. He received the B.Sc. degree in mechanical engineering-thermo-fluids from Shiraz University, Shiraz, Iran, the M.S. degrees in mechanical systems and solid mechanics from Shiraz University and the University of Central Florida, Orlando, FL, USA, and the Ph.D. degree in mechanical engineering from Drexel University, Philadelphia, in 2015. 
\end{IEEEbiography}

\end{document}